\documentstyle[12pt,a4]{article}
\begin{document}
\hyphenation{anti-fermion}
\baselineskip = 7mm
\begin{center}
\begin{Large}
{\bf {  Hyperfine Anomaly of Be Isotopes and

Anomalous Large Anomaly in $^{11}$Be }}
\end{Large}
\vspace{1cm}

T. FUJITA\footnote{e-mail: fffujita@phys.cst.nihon-u.ac.jp} and
K. ITO\footnote{e-mail: kito@shotgun.phys.cst.nihon-u.ac.jp}

Department of Physics, Faculty of Science and Technology

Nihon University, Tokyo, Japan

and

Toshio SUZUKI\footnote{e-mail:suzuki@chs.nihon-u.ac.jp}

Department of Physics, College of Humanities and Sciences, Nihon 
University
Sakurajosui 3-25-40, Setagaya-ku, Tokyo 156-8550, Japan
\vspace{2cm}

{\large ABSTRACT}

\end{center}
A new result of investigations of the hyperfine structure (hfs) anomaly
in Be isotopes is presented.  The hfs constant for $^{11}$Be is obtained
by using the core plus neutron type wave function:
$ |2s_{1\over 2}>+|1d_{5\over2}\times 2^+ ; \frac{1}{2}^{+}> $.
A large hfs anomaly of $^{11}$Be is found, which is mainly
 due to a large radius of the halo single particle state.

\vspace{2cm}
\noindent
$PACS \  numbers $ : 21.10.Ky, 21.60.Cs, 27.20.+n, 33.15.Pw \\
$Key \ words $ : Hyperfine anomaly, neutron halo

\newpage
\section{Introduction}

Recently, much interest has been paid to
the magnetic hyperfine structure (hfs)
for various nuclear isotopes
\cite{Kla,Fin,Sch,Asa,AF,AFH}.
For the experimental side,
there is some progress in observing
the transition of the hyperfine levels for atomic ground states.
In fact, the accuracy of the measurement of the hfs splitting
is improved a great deal, and
even for lighter nuclei, there is a good chance
of observing the hfs anomaly. This becomes
possible due to the ion trap method which can isolate the atoms.
This ion trap method\cite{End,Wada} can measure the hfs anomaly
with the accuracy of  order $10^{-6}$.

The hyperfine structure has a sensitivity to the magnetization distribution 

in nucleus.  It can, therefore, present a unique way to measure
the neutron distribution in nucleus.
In particular, there is a strong evidence that the $^{11}$Li has a neutron 

halo\cite{Tan} which may extend quite far over the typical nuclear radius
of the neighboring light nuclei.
Moreover, $^{11}$Be has an anomalous spin-parity state for the ground
state\cite{FOT}, and there may be some chance that it also has a large 
neutron
radius.

In this paper, we present a model calculation of the hfs anomaly
for Be isotopes. For $^{7}$Be and $^{9}$Be nuclei, we can use the 
Cohen-Kurath
wave functions\cite{CK}.  On the other hand, $^{11}$Be has an anomalous
spin-parity state, which is ${1\over 2}^+$ instead of ${1\over 2}^-$.
A simple minded shell model wave function does not
give a proper structure of the ground state.

Here, we calculate the matrix element of the hfs operator
for the ground state of $^{11}$Be using
the core plus neutron type wave function:
$ \alpha |2s_{1\over 2}>+ \beta |1d_{5\over2}\times 2^+: {1\over2}^+> $.
It turns out that the hfs anomaly $\epsilon$ for $^{11}$Be  is quite large 

in magnitude compared to those of $^{7}$Be and $^{9}$Be.
This is mainly related to the fact that the ground
state of $^{11}$Be has a large
radius compared to those of ground states of other Be isotopes.

The paper is organized as follows. In the next section, we briefly
explain the theory of the magnetic hyperfine structure
in electronic atoms. Section 3 treats the core plus neutron
type wave function. We evaluate the matrix elements
of the hfs operator using this wave function.
In section 4,  numerical results of the isotope
shifts of the hfs anomaly for Be isotopes are presented.
Summary is given in section 5.


\section{Magnetic Hyperfine Structure}
An atomic electron which is bound by a nucleus feels the magnetic 
interaction
in addition to the static Coulomb force.
The magnetic interaction between the electron and the nucleus
can be described as
$$ H' = -\int {\bf j}_N ({\bf r}) \cdot {\bf A}({\bf r}) d^3 r
\eqno{(2.1)} $$
where the nuclear current ${\bf j}_N ({\bf r})$ can be written as
$$ {\bf j}_N({\bf r}) = {e\hbar\over{2Mc}}\sum_i g_s^{(i)} \nabla \times s_i 
\delta ({\bf r}-{\bf R}_i )
+\sum_i  {eg^{(i)}_{\ell}\over{2M}} \left( {\bf P}_i \delta ({\bf r}-{\bf 
R}_i )
+\delta ({\bf r}-{\bf R}_i ) {\bf P}_i \right) \eqno{(2.2)}  $$
${\bf A}({\bf r})$ denotes the vector potential which is created by the
atomic electron, and it can be written as
$$ {\bf A}({\bf r})= \int { {\bf j}_e({\bf r}')
\over{|{\bf r}-{\bf r}' |}} d^3 r' \eqno{(2.3)} $$
where ${\bf j}_e ({\bf r}) $ denotes the current density of an electron,
and can be written as
$$ {\bf j}_e ({\bf r}) = -e \mbox{\boldmath $\alpha $}
\delta ({\bf r}-{\bf r}_e )
\eqno{(2.4)} $$
where $\mbox{\boldmath $\alpha $} $ denotes the Dirac matrix.

\vspace{1cm}
\subsection{The hfs anomaly}

The magnetic hyperfine splitting
energy $W$ can be written as
$$W = <IJ:FF|H'|IJ:FF>= {1\over 2} \left[ F(F+1)-I(I+1)-J(J+1) \right] a_I 
\eqno{(2.5)} $$
where $I$,$J$ and $F$ denote the  spin of the nucleus, the spin of the 
atomic
electron and the total spin of the atomic system, respectively. $a_I$ is 
called
the hyperfine structure (hfs) constant.
Following ref. \cite{FA}, we can write the expression for the $a_I$ as
$$ a_I =a_I^{(0)} (1+\epsilon ) \eqno{(2.6)} $$
where $a_I^{(0)}$ is the hfs constant for the point charge, and
can be written as
$$ a_I^{(0)} = -{2ek\mu_N \over{IJ(J+1)}}
\mu  \int_0^\infty F^{(kJ)} G^{(kJ)} dr \eqno{(2.7)} $$
where $\mu$ is the magnetic moment of the nucleus in units of the nuclear
magneton $\mu_{N}$, and $F^{(kJ)}$ and $ G^{(kJ)}$ are the large and small 

components of the relativistic electron wave function for the $kJ$ state.
$\epsilon$ is called hfs anomaly and can be written as
$$\epsilon =-{0.62 b^{(kJ)}\over{\mu}}<II|\sum_{i=1}^A
\left( {R_i\over{R_0}}\right)^2
\mu_i |II>
-{0.38 b^{(kJ)}\over{\mu}}<II|\sum_{i=1}^A
\left( {R_i\over{R_0}}\right)^2  g_s^{(i)}
\Sigma_i^{(1)} |II> \eqno{(2.8)} $$
where $\mu_{i}$ is the single-particle operator of the magnetic
moment, that is,
$\mu_{i} = g_{s}^{(i)}s_{i}$ + $g_{\ell}^{(i)}\ell_{i}$, and
$ \Sigma_i^{(1)}$ is defined as
$$ \Sigma_i^{(1)} = s_i +\sqrt{2\pi}[sY^{(2)}]_i^{(1)} . \eqno{(2.9)} $$
$R_0$ is a nuclear radius and can be given as
$ R_0=r_0 A^{ 1\over 3} $ with $r_0 = 1.2$ fm.
On the other hand, $b^{(kJ)}$ is a constant
which is calculated in terms of relativistic electron
wave functions\cite{BW}, and is given as
$$ b^{(kJ)} = 0.23 k^2_0 R_0 \gamma (1-0.2 \gamma^2 )
/ \int_0^{\infty} F^{(kJ)}G^{(kJ)} dr . \eqno{(2.10)} $$
$m_e$ denotes the electron mass, $k_0^2$ is a normalization constant,
and $\gamma = Z\alpha $.

The isotope shift of the hfs anomalies of the two isotopes
$\Delta_{12}$ is defined as
$$ \Delta_{12} = {a_{I_1}g_2\over{a_{I_2}g_1}} -1  \eqno{(2.11)} $$
where $g_{1}$ and $g_{2}$ are the total nuclear g-factors for isotope 1 and 
2,
respectively.
Since the hfs anomaly $\epsilon$ is quite small, $\Delta_{12}$ becomes
$$ \Delta_{12} \approx \epsilon_1 -\epsilon_2 . \eqno{(2.12)} $$

The hfs anomaly $\epsilon$ can be calculated once we know the nuclear
wave function.
Here, we employ  simple-minded shell model wave functions with
core polarization taken into account. We take
the following approach which considers
only the $\Delta \ell =0$ core polarization for the $ \Sigma_i^{(1)} $
operator. In this case, we can obtain the matrix element of the $ 
\Sigma_i^{(1)}$
 without introducing any free parameters as discussed in ref. \cite{FA}.

\vspace{1cm}
\subsection{$\Delta \ell =0$ core polarization }

In this case, we can express the effect of the core polarization
on the $ \Sigma_i^{(1)} $ operator in terms of the core polarization
of the magnetic moment. Following ref. \cite{FA}, we can write the 
expectation
value of the $ \Sigma_i^{(1)} $ as
$$ <II| \sum_{i=1}^A g_s^{(i)}  \Sigma_i^{(1)} |II>
= \pm g_s^{(VN)}{3(I+{1\over 2})\over{4(I+1)}} +
{3g_s^{(VN)}\over{4(g_s-g_{\ell})^{(VN)} }} (\mu-\mu_{sp}-
\delta \mu^{mes} ) \eqno{(2.13)} $$
for $I=\ell \pm {1\over 2} $.
Here, $g_s^{(VN)}$ denotes the g-factor of the valence nucleon
for the single particle state we are considering. $\mu_{sp}$ is the single 

particle value of the magnetic moment.
$\delta \mu^{mes}$
comes from the
meson exchange current and can be given approximately as
$$ \delta \mu^{mes} \approx  0.1 \tau_3 \ell   . $$
Therefore, we do not have any free parameters in the evaluation of
the expectation value of the $ \Sigma_i^{(1)} $.  As the Be isotopes have
orbits with small $\ell $, the exchange current effects are not important 
and
we can safely neglect the term $\delta \mu^{mes} $.
\vspace{2cm}
\section{Core plus neutron model}

Recently, Suzuki et al.\cite{SOM} investigated the magnetic moment of
the ground state of $^{11}$Be. They describe the ground state
of  $^{11}$Be as
$$ |^{11}\mbox{Be} \left({1\over {2}}^+\right)> = \alpha |^{10}\mbox{Be}
(0^+) \times \nu
2s_{1\over 2}: {1\over 2}^+ > + \beta  |^{10}\mbox{Be} (2^+) \times \nu
2d_{5\over 2}: {1\over 2}^+ >  .  \eqno{(3.1)} $$
In this case, the magnetic moment of the $^{11}$Be ground state can be
expressed  as
$$ \mu = \alpha^2 \mu_{\nu (2s_{1\over 2}) } +{7\over {15}} \beta^2
\mu_{\nu (1d_{5\over 2})} -{1\over 3} \beta^2 \mu_{(2^+)}  \eqno{(3.2)} $$
where $\mu_{\nu (2s_{1\over 2}) }$, $\mu_{\nu (1d_{5\over 2}) }$ and
$\mu_{(2^+)}$ denote the magnetic moment of the neutron
$ 2s_{1\over 2}$, $ 1d_{5\over 2}$ states
and the $2^+$ (3.37 MeV) state of $^{10}$Be, respectively.
Here, we evaluate the hfs anomaly $\epsilon$ for the two- component wave
function, eq.(3.1).
The hfs anomaly of eq.(2.8) can be written as
$$ \epsilon =
-{0.62 b^{(1s)}\over{\mu}} \left[
\alpha^2 \mu_{\nu (2s_{1\over 2})} < \left( {R\over{R_0}} \right)^2 >_{2s} + 
{7\over {15}} \beta^2
\mu_{\nu (1d_{5\over 2})} < \left( {R\over{R_0}} \right)^2 >_{1d}
-{1\over 3} \beta^2 \mu_{ (2^+ )} < \left( {R\over{R_0}} \right)^2 >_{1p}
\right] $$
$$ -{0.38 b^{(1s)}\over{\mu}} \left[
\alpha^2 \Sigma_{\nu (2s_{1\over 2})} < \left( {R\over{R_0}} \right)^2 
>_{2s} +
{7\over {15}} \beta^2
\Sigma_{\nu (1d_{5\over 2})}  < \left( {R\over{R_0}} \right)^2 >_{1d}
-{1\over 3} \beta^2 \Sigma_{ (2^+ )} < \left( {R\over{R_0}} \right)^2 >_{1p} 
\right] \eqno{(3.3)} $$
where
$$ \Sigma_{j} = < j j \mid \sum_{i} g_{s}^{(i)} \Sigma_{i}^{(1)} \mid jj > 
\eqno{(3.4)} $$
 and $i$ runs over the nucleons in the state that has
 the total angular momentum $j$.
$\mu_{\nu(2s_{{1\over 2}})}$ = $\mu_{\nu(1d_{{5\over 2}})}$ =
${1\over 2} g_{s}^{(n)}$, where $g_{s}^{(n)} = -3.826$ is
the spin g-factor for the neutron, and $\mu_{(2^{+})}$ = 1.787\cite{SOM}.
The values of
$\Sigma_{\nu (2s_{1\over 2})}$, $\Sigma_{\nu (1d_{5\over 2})}$ and
$\Sigma_{(2^{+})}$ are obtained as   $ \Sigma_{\nu (2s_{1\over 2})}$ =
${1\over 2} g_{s}^{(n)}$, $ \Sigma_{\nu (1d_{5\over 2})}$ =
${9\over 14} g_{s}^{(n)}$, and  $ \Sigma_{(2^{+})} = -1.034$, respectively. 

The $\mu_{(2^{+})}$ and the $\Sigma_{(2^+)}$ are obtained by using
the Cohen-Kurath wave function, TBE(8-16)\cite{CK,OXB}.
Now, the recent measurement of the magnetic moment of $^{11}$Be
gives\cite{Asah}
$$ \mu ({^{11}Be}) = -1.682(3) \ \ \mu_N  .  \eqno{(3.5)}$$
This indicates that the value of $\alpha^2$ is close to
$$ \alpha^{2} = 0.5  \eqno{(3.6)}$$
from the comparison of the observation with Fig. 2a of ref. \cite{SOM}.
In this way, we obtain the hfs anomaly $\epsilon$ for $^{11}$Be.
For the $^{7}$Be and $^{9}$Be isotopes, we can calculate the hfs anomaly
using the Cohen-Kurath wave function\cite{CK}.
The hfs anomaly $\epsilon$ for this case is written similarly as
$$ \epsilon = - {0.62 b^{(1s)}\over \mu} \mu_{({3\over 2}^{-})}
<({R \over R_{0}})^{2}>_{1p} - {0.38 b^{(1s)}\over \mu}
\Sigma_{({3\over 2}^{-})}
<({R \over R_{0}})^{2}>_{1p}. \eqno{(3.7)}$$

We also consider the case in which the effective $g$ factor for the  spin
operator, $g_{s}^{eff}$, is taken into account.  As for the halo orbit,
we assume that the second-order effects are rather small and result
in little quenching of the spin operator.  We, therefore,
take $g_{s}^{eff}$ = 1.0 for the $\nu$2s$_{1/2}$- orbit.
In this case,
$\mu_{\nu(2s_{1/2})}$ = $\frac{1}{2}g_{s}^{(n)}$ = $-1.913$,
$\mu_{\nu(1d_{5/2})}$ = $\frac{1}{2}g_{s}^{(n)} g_{s}^{eff}$
= $-1.913 g_{s}^{eff}$,
$\mu_{(2^{+})}$ = 1.076 + 0.711 $g_{s}^{eff}$,
and
$\Sigma_{\nu(2s_{1/2})}$ = $\frac{1}{2}g_{s}^{(n)}$ = $-1.913$,
$\Sigma_{\nu(1d_{5/2})}$ = $\frac{9}{14}g_{s}^{(n)} g_{s}^{eff}$
= $-2.460 g_{s}^{eff}$,
$\Sigma_{(2^{+})}$ = $-1.034 g_{s}^{eff}$,
in units of $\mu_{N}$.
The magnetic moment is, then, given by
$$ \mu = -1.913 \alpha^{2} - (0.3585 + 1.1298 g_{s}^{eff})\beta^{2}.
\eqno{(3.8)}$$

In comparison with the experimental value of the magnetic moment,
$-1.682(3)$ $\mu_{N}$, we obtain $\alpha^{2}$ = 61$\%$.
When the contributions from the meson exchnge currents are taken into 
account,
$\alpha^{2}$ is close to 65$\%$.  We give results of the hfs anomaly
$\epsilon$ for $^{11}$Be also for the case with the effective spin $g$ 
factor.
Numerical evaluations are carried out in section 4.

\vspace{2cm}
\section{Numerical Results}

Now, we evaluate numerically the hfs anomaly $\epsilon$ for the Be isotopes. 

\subsection{Core plus neutron type wave function}
In order to evaluate eq.(3.3), we need to know the magnetic
moment $\mu$ for the Be isotopes. The magnetic moments of
$^9$Be and $^{11}$Be
are observed, but the magnetic moment of $^7$Be
is not yet determined experimentally.
We evaluate the magnetic moment of $^{7}$Be empirically.

For $^7$Be, we can make use of the magnetic moment of $^7$Li
since they are isospin doublet states. In this case, we predict
the magnetic moment for $^7$Be
$$ \mu^{Pred} (^7\mbox{Be}) = -1.377 . \eqno{(4.1)}$$

In Table 1, we list the values of the quantities which are necessary
to calculate the hfs anomaly $\epsilon$.
The r.m.s. radii are those for wave functions solved in a Woods-Saxon 
potential
with $R_{0}$ = 1.2 A$^{1/3}$ and a=0.60 fm.  For the 2s$_{1/2}$ and 
1d$_{5/2}$
neutron orbits in $^{11}$Be, the wave functions are obtained to reproduce
the separation energies, 0.50 MeV and 3.87 MeV, respectively.
The anomaly $\epsilon$ for $^{11}$Be is obtained from eq. (3.3) to be
$$ \epsilon (^{11}\mbox{Be}) = -0.12015 \alpha^{2} - 0.02326 \beta^{2}
\quad ( \% ). \eqno{(4.2)}$$
 It gives $\epsilon = -0.0717 \%$ for $\alpha^{2}$ = 0.50 and
 $\beta^{2}$ = 0.50.  Magnitude of $\epsilon$ gets as large as
 $-0.091 \%$ ($-0.101 \%$) as $\alpha^{2}$ becomes 0.70 (0.80).
The hfs anomalies for $^{7}$Be and $^{9}$Be are
obtained from eq.(3.7) by using the Cohen-Kurath
wave functions\cite{CK,OXB}.

In Table 2a, we present the calculated values of hfs anomay $\epsilon$
for the Be isotopes.
Now, it turns out that the hfs anomaly for $^{11}$Be has a very large value 

compared to other isotopes. This is mainly connected with the fact that the 

$^{11}$Be has a large neutron radius.
The r.m.s. radius of the halo 2s$_{{1\over
2}}$ orbit becomes as large as 6.4 fm in deformed Woods-Saxon 
models\cite{Muta}.
When we use a value of 6.37 fm for the r.m.s. radius of the halo
2s$_{{1\over
2}}$ orbit, which is obtained in the deformed Woods-Saxon
model\cite{Muta,SOM},
the hfs anomaly is given by
$$ \epsilon = -0.12827 \alpha^{2} -0.02326 \beta^{2} \quad (\%).
\eqno{(4.3)}$$
Eq. (4.3) leads to $\epsilon = -0.0758$ $\%$ for $\alpha^{2}$ = 0.5,
which is close to the value obtained from eq. (4.2).
The anomaly $\epsilon$ becomes $-0.076 \sim -0.097$ $\%$
for $\alpha^{2}$ = $0.5 \sim 0.7$.
We finally give numerical results for the case with the effective spin
$g$ factor.  We take $g_{s}^{eff}$ = 1.0 for the halo $\nu$2s$_{1/2}$ orbit 
and
$g_{s}^{eff}$ = 0.85\cite{CWB} for other normal orbits.
The hfs anomaly is given by
$$ \epsilon = -0.12015 \alpha^{2} - 0.02014 \beta^{2} \quad(\%).
\eqno{(4.4)}$$
Here, the 2s$_{1/2}$ orbit obtained in the spherical Woods-Saxon potential 

is used.  The anomaly becomes $\epsilon$ = $-0.0852 \%$ for $\alpha^{2}$ = 
0.65,
whose magnitude is larger than the value obtained from eq. (4.2) about
by 0.01 $\%$.  The anomaly $\epsilon$ becomes $-0.070 \sim -0.100 \%$ for
$\alpha^{2}$ = 0.5$\sim$ 0.7.
\vspace{1cm}
\subsection{Single particle model with core polarization}
Next, we calculate the hfs anomaly $\epsilon$ for Be isotopes using the 
single
particle model with core polarization.
Since the radius of the single particle state can be different from that of 

the core polarization state, we modify the expression eq. (2.8) in the same 
way
as eq. (3.3)
$$ \epsilon =
-{0.62 b^{(1s)}\over{\mu}} \left[
\mu_{sp} < \left( {R\over{R_0}} \right)^2 >_{sp} +
\delta\mu_{CP} < \left( {R\over{R_0}} \right)^2 >_{CP}
\right] $$
$$ -{0.38 b^{(1s)}\over{\mu}} \left[
g_{s}^{(VN)}\Sigma_{sp} < \left( {R\over{R_0}} \right)^2 >_{sp} +
{3\over 4}\delta\mu_{CP} < \left( {R\over{R_0}} \right)^2 >_{CP}
\right] \eqno{(4.5)} $$
 where $\Sigma_{sp}$ denotes the expectation value of
$ <II \mid \sum_{i=1}^{A} \Sigma_{i}^{(1)} \mid II>$ with the single 
particle
state and can be written as
$$ \Sigma_{sp} = {3(I+{1\over2})\over{4(I+1)}}
\quad \mbox{for} \quad I=\ell + {1\over2}. \eqno{(4.6)}$$
 $\delta\mu_{CP}$ is
the magnetic moment which arizes from the core polarization\cite{NAH}.
$< ({R\over{R_{0}}})^{2} >_{CP}$ denotes the expectation values with the 
states
involved in the core polarization.  In Be isotopes, they are 1p states.
In Table 2b, we list the calculated values of the hfs anomaly $\epsilon$
for the Be isotopes.
In the same way as the core plus neutron type calculation,
the hfs anomaly for $^{11}$Be has a very large
value compared to other isotopes. This is due to the fact
that the $^{11}$Be has a large neutron radius since it has an anomalous
spin-parity state.
\vspace{3cm}
\section{Conclusions}

We have presented the numerical calculations of
the magnetic hfs anomaly for the Be isotopes.
First, we employ the wave function which has a component coupled to the
$2^+-$core excitation.
This gives a large hfs anomaly for $^{11}$Be since the neutron outside
the shell is assumed to be $2s_{1\over 2}$ or $1d_{5\over 2}$ orbits.
On the other hand,  we find that the $^{7}$Be and $^{9}$Be isotopes have
a small hfs anomaly.
We also evaluate the hfs anomaly
using a single particle shell model with core polarization.
We also predict a very large hfs anomaly
for $^{11}$Be.

It would be extremely interesting to learn
whether the very large hfs anomaly of $^{11}$Be
can be realized in nature or not.  Since this is related to the radius of
the neutron halo nucleus, it may well help understand the structure of
the neutron-rich nuclei.
We hope that experimental observations will clarify this point in near 
future.

\vspace{3cm}

\section*{Acknowledgments}
The authors would like to thank  I. Katayama and M. Wada for discussions.
They are also grateful to K. Asahi for discussions  on the recent 
measurement
of the magnetic moment of $^{11}$Be.  They also thank T. Otsuka for valuable 

discussions on the magnetic moment of $^{11}$Be.
This work is supported in part by Japanese-German Cooperative Science
Promotion Program and Grant-in-Aids for Scientific Research (c) (No. 
08640390)
from the Ministry of Education, Science, Sports and Culture.

\newpage



\newpage
\begin{center}
\underline{Table 1} \\
$\displaystyle{ \rm  \  R_0, \  b^{(1s)}, \
r.m.s. \  radius \ and \  magnetic \  moment  }  $ \\
\ \ \ \\
\begin{tabular}{|c|c|c|c|}
\hline
\  & \  & \ & \  \\
\  & $^7\mbox{Be} $ & $^9\mbox{Be} $  & $ ^{11}\mbox{Be} $   \\
\hline
\hline
$ R_0$ [fm]  & 2.296 & 2.496 & 2.669 \\
\hline
$ b^{(1s)} $ [\%] & 0.0170 & 0.0185 & 0.0198 \\
\hline
$<r^2>^{1\over 2}_{(1p)}$ [fm] & 2.553 & 2.569 & 2.588 \\
\hline
$<r^2>^{1\over 2}_{(2s)}$ [fm] & $--$ & $--$ & 6.165 \\
\hline
$<r^2>^{1\over 2}_{(1d)}$ [fm] & $--$ & $--$ & 3.551 \\
\hline
$\mu_{exp}$  [n.m.] & $-1.377^*$ & $ -1.177$ & $-1.682$ \\
\hline
$\mu_{sp}$  [n.m.] & $-1.913$ & $-1.913$ & $-1.913$ \\
\hline
$\delta \mu_{CP}$  [n.m.] & $0.536^*$ & 0.736 & 0.231 \\
\hline
\end{tabular}

\vspace{0.7cm}
\begin{minipage}{13cm}
The values of $R_0$, $b^{(1s)}$, r.m.s. radii for the
$1p$ and $2s$ states and the magnetic
moments for Be isotopes are shown.
* indicates that the value of the magnetic moment
is empirically extracted from those of other
nuclear isotopes.
\end{minipage}
\end{center}

\vspace{1cm}
\begin{center}
\underline{Table 2a} \\
$\displaystyle{ \rm \  hfs \  anomaly \
\epsilon \  and \  isotope \  shift \ \Delta_{12}   }  $ \\
\ \ \ \\
\begin{tabular}{|c|c|c|c|}
\hline
\  & \  & \ & \  \\
\  & $^7\mbox{Be} $ & $^9\mbox{Be} $  & $ ^{11}\mbox{Be} $   \\
\hline
\hline
$\epsilon$ [\%] & $-0.0245$  & $-0.0249$  & $-0.0717$   \\
\hline
$\Delta_{7,A}$ [\%] & 0 & $$0.0004 & 0.0472  \\
\hline
\end{tabular}

\vspace{1cm}
\begin{minipage}{13cm}
The calculated values of the hfs anomaly $\epsilon$ and the isotope shift 
$\Delta_{12}$ for the Be isotopes obtained from eqs. (3.3) and (3.7) are 
shown.
$\alpha^{2}$ = $\beta^{2}$ = 0.50 are used for
$^{11}$Be.
\end{minipage}
\end{center}

\vspace{1cm}
\begin{center}
\underline{Table 2b} \\
$\displaystyle{ \rm \  hfs \  anomaly \
\epsilon \  and \  isotope \  shift \ \Delta_{12}   }  $ \\
\ \ \ \\
\begin{tabular}{|c|c|c|c|}
\hline
\  & \  & \ & \  \\
\  & $^7\mbox{Be} $ & $^9\mbox{Be} $  & $ ^{11}\mbox{Be} $   \\
\hline
\hline
$\epsilon$ [\%] & $-0.024$  & $-0.023$  & $-0.118 $   \\
\hline
$\Delta_{7,A}$ [\%] & 0 & $-$0.001 & 0.094  \\
\hline
\end{tabular}

\vspace{1cm}
\begin{minipage}{13cm}
The calculated values of the hfs anomaly $\epsilon$ and the isotope shift 
$\Delta_{12}$ for the Be isotopes obtained from eq. (4.5) are shown.
\end{minipage}
\end{center}

\end{document}